\documentclass[a4paper,fleqn,usenatbib]{mnras}

\usepackage[T1]{fontenc}
\usepackage{ae,aecompl}

\usepackage{ulem}

\usepackage{graphicx}	% Including figure files
\usepackage{amsmath}	% Advanced maths commands
\usepackage{amssymb}	% Extra maths symbols    

\usepackage{natbib}
\usepackage{mathtools}

\usepackage[dvipsnames]{xcolor}

\def\gtsim {>\kern-1.2em\lower1.1ex\hbox{$\sim$}~}   
\def\ltsim {<\kern-1.2em\lower1.1ex\hbox{$\sim$}~}   

\usepackage{newtxtext,newtxmath}

\title[Nucleosynthesis signatures of proto-NS winds]{Nucleosynthesis signatures of neutrino-driven winds from proto-neutron stars: a perspective from chemical evolution models}

\author[F. Vincenzo et al.]{Fiorenzo Vincenzo$^{1}$\thanks{email: vincenzo.3@osu.edu}, 
Todd A. Thompson$^{1}$, David H. Weinberg$^{1,2}$, Emily J. Griffith$^{1}$, 
\newauthor 
James W. Johnson$^{1}$, Jennifer A. Johnson$^{1}$
\\ ~ \\
$^{1}$Department of Astronomy \& Center for Cosmology and AstroParticle Physics, The Ohio State University, Columbus, OH 43210, USA \\
$^{2}$Institute for Advanced Study, Princeton, NJ 08540, USA
}

\begin{document}

\date{Accepted 2021 September 27. Received 2021 September 06; in original form 2021 February 09}

\pagerange{\pageref{firstpage}--\pageref{lastpage}} \pubyear{2021}

\maketitle

\label{firstpage}

%%%%%%%%%%%%%%%%%%%%%%%%%%%%%%%%%%%%%%
%%%%%%%%%%%%%%%%%%%%%%%%%%%%%%%%%%%%%%

\begin{abstract}
We test the hypothesis that the observed first-peak (Sr, Y, Zr) and
second-peak (Ba) s-process elemental abundances in low-metallicity Milky Way stars, and the abundances of the elements
Mo and Ru, can be explained by a pervasive r-process contribution originating in neutrino-driven winds from highly-magnetic and rapidly rotating proto-neutron stars (proto-NSs). We construct chemical evolution models that incorporate recent calculations of proto-NS yields in addition to contributions from AGB stars, Type Ia supernovae, and two alternative sets of yields for massive star winds and core-collapse supernovae. For non-rotating massive star yields from either set, models without proto-NS winds underpredict the observed s-process peak abundances by $0.3$-$1\,\text{dex}$ at low metallicity, and they severely underpredict Mo and Ru at all metallicities.  Models incorporating wind yields from proto-NSs 
with spin periods $P \sim 2$-$5\,\text{ms}$ fit the observed trends
for all these elements well.  Alternatively, models omitting proto-NS winds but adopting yields of rapidly rotating massive stars, with $v_{\rm rot}$ between $150$ and $300\,\text{km}\,\text{s}^{-1}$, can explain the observed abundance levels reasonably well for $\text{[Fe/H]}<-2$. These models overpredict [Sr/Fe] and [Mo/Fe] at higher metallicities, but with a tuned dependence of $v_{\rm rot}$ on stellar metallicity they might achieve an acceptable fit at all [Fe/H]. If many proto-NSs are born with strong magnetic fields and short spin periods, then their neutrino-driven winds provide a natural source for Sr, Y, Zr, Mo, Ru, and Ba in low-metallicity stellar populations. Conversely, spherical winds from unmagnetized proto-NSs overproduce the observed Sr, Y, and Zr abundances by a large factor.
\end{abstract}

%%%%%%%%%%%%%%%%%%%%%%%%%%%%%%%%%%%%%%
%%%%%%%%%%%%%%%%%%%%%%%%%%%%%%%%%%%%%%


 %%%%%%%%%%%%%%%%%%%%%%%%%%%%%%%%%%%%%%
 %%%%%%%%%%%%%%%%%%%%%%%%%%%%%%%%%%%%%%

\section{Introduction} \label{sec:intro}

Isotopes with nucleon number larger than that of iron-peak elements ($\mathcal{A}\approx 56$) are prevented by an increasingly large Coulomb barrier from being synthesized by any charged-particle-induced thermonuclear reaction in stellar interiors. The existence of the vast majority of the trans-iron elements is explained by a series of neutron-capture processes followed by $\beta$-decays, starting from seed nuclei of iron-peak elements \citep{burbidge1957,cameron1957}.
Neutron-capture nucleosynthesis can operate by means of two processes, depending on whether the rate of neutron-capture is slow (s-process) or  rapid (r-process) with respect to the rate of $\beta$-decay. Nearly all trans-iron elements are produced by a mixture of s- and r-process events, except for a number of stable s-only nuclei which are shielded from any r-process contribution by the presence of a stable, neutron-rich r-only nucleus with equal $\mathcal{A}$.
 
In the classical picture, the s-process proceeds along the so-called valley of $\beta$-stability, giving rise to three distinctive peaks in the Solar abundance template, which correspond to the magic neutron numbers $\mathcal{N}_{n} = 50$ (e.g., $^{88}$Sr, $^{89}$Y, $^{90}$Zr), $\mathcal{N}_{n} = 82$ (e.g., $^{138}$Ba), and $\mathcal{N}_{n} = 126$ (e.g., $^{208}$Pb). 
Such peaks in the Solar abundance pattern are due to the fact that the stable heavy-element isotopes with a magic neutron number have a neutron-capture cross-section which is much lower than that of the other isotopes of the $\beta$-stability valley, creating a bottleneck in the s-process path. 

The main site of the s-process nucleosynthesis in astrophysical environments is found in the late evolutionary stages of low- and intermediate-mass (LIM) stars, during the asymptotic giant branch (AGB) phase (e.g., \citealt{ulrich1973}). The main source of free-neutrons in AGB stars is provided by the reaction $^{13}$C($\alpha$,$n$)$^{16}$O \citep{iben1982,hollowell1989}, which is active at energies $E \approx 8 \, \text{keV}$. The formation of the so-called $^{13}$C-pocket in the He intershell during the interpulse period of AGB stars represents the key physical mechanism responsible for the s-process nucleosynthesis in LIM stars \citep{iben1982,hollowell1989,straniero1995,gallino1998,straniero2009}. The continuous recurrence of the third-dredge up and interpulse periods makes s-process nucleosynthesis in AGB stars very effective for a relatively extended period of time, producing remarkable effects in the chemical abundance distribution observed in the stars of our Galaxy (see, for example, \citealt{busso2001,karakas2007,cristallo2009,karakas2016}). 

S-process nucleosynthesis can also take place during the evolutionary stage of core He-burning of massive stars, which -- before exploding as core-collapse supernovae (SNe) -- can pollute the interstellar medium (ISM) of our Galaxy through radiatively-driven stellar winds, enriched with He, C, N, O, and traces of s-process elements. In particular, massive stars provide a prompt s-process contribution to the chemical evolution of our Galaxy at low metallicities, before low-mass stars reach the AGB phase and enrich the ISM with their nucleosynthetic products. 
An interesting study is that of \citet{cescutti2013}, who proposed that a prompt s-process contribution from rapidly-rotating massive stars \citep{frischknecht2012} -- predicted to be more abundant at low metallicities, where stars are more compact and hence rotate faster -- can help explain the shape of the scatter in [Sr/Ba] in metal-poor halo stars (see also the review of \citealt{frebel2010}). 

The r-process works by rapidly piling up neutrons in an increasingly heavy nucleus. As the neutron-capture proceeds, the nucleon binding energy becomes increasingly low, until an equilibrium is reached between the photo-disintegration and the neutron-capture, which temporarily freezes out the neutron number. The $\beta$-decay of the neutron-rich nucleus then breaks the deadlock, allowing the r-process to resume. It then proceeds rapidly until a new equilibrium is reached again (see, for example, the classical books of \citealt{clayton1983,rolfs1988}). 

The r-process is the main physical mechanism behind the nucleosynthesis of the neutron-rich heavy elements observed in Milky Way (MW) stars at all metallicities, but it also provides a non-negligible contribution to the majority of neutron-capture elements which are observed in metal-poor stars. The signature of r-process events is very pervasive, and it is observed at all metallicities in the stars of our Galaxy.  

The astrophysical environments of r-process events in the cosmos usually involve explosive physical conditions, because very high neutron densities are required. The neutron-rich isotopes produced along the r-process path have a very short half-life, causing the neutron-rich nuclei to beta-decay over timescales of the order of milliseconds, if they did not undergo further rapid neutron-capture. 
Examples of r-process sites that have been proposed and investigated by theoretical studies include 
\textit{(i)} neutrino-driven winds from proto-neutron stars (NSs) 
\citep{woosley1994,takahashi1994, qian1996,hoffman1997,otsuki2000,thompson2001,wanajo2009,hansen2013,wanajo2013}; \textit{(ii)} neutrino-driven winds from highly magnetic and potentially rapidly rotating proto-magnetars \citep{thompson2003,thompson2004,metzger2007,metzger2008,vlasov2014,vlasov2017,thompson2018};
\textit(iii) neutrino-driven winds around the accretion disk of a black hole \citep{pruet2004,metzger2008,wanajo2012,siegel2019}; \textit{(iv)} electron-capture SNe (see, for example, \citealt{wanajo2009,cescutti2013,kobayashi2020}); \textit{(v)} magneto-rotationally-driven SNe \citep{burrows2007,winteler2012,nishimura2015,nishimura2017,mosta2014,mosta2015,mosta2018,cescutti2014,halevi2018,reichert2021}; and \textit{(vi)} neutron-star mergers \citep{lattimer1977,freiburghaus1999,agarst2004,goriely2011,rosswog2013,matteucci2014,cescutti2015,vincenzo2015,kobayashi2020}. Since it is likely that all these mechanisms have contributed to r-process nucleosynthesis at some level, the theoretical studies have focused on exploring the frequency of each event and the predicted template of the corresponding r-process ejecta.

Due to the large uncertainties in the r-process nucleosynthesis calculations, the working strategy of the first MW chemical evolution models was to assume empirical yields for the r-process, which were tuned to reproduce the abundances of light and heavy neutron-capture elements at low metallicity. After doing this, it was possible to discuss the conditions for the different r-process sites to produce enough material to explain the neutron-capture elemental abundances and their dispersion in metal-poor stars (e.g., \citealt{travaglio2004,cescutti2006,cescutti2008}). Another strategy was employed by \citet{prantzos2018}, who emphasize the rotating massive star contribution and fit the metallicity dependence of stellar rotation empirically by reproducing observed abundance trends (see also \citealt{prantzos2020}). \citet{kobayashi2020} recently explored the impact of a variety of different scenarios to the observed abundances of neutron-capture elements.

In this work we explore the hypothesis that neutrino-driven winds from proto-NSs can explain the abundances of the first peak (Sr, Y, Zr) and second peak (Ba) of the s-process elemental abundance distribution. We test this hypothesis by assuming that massive stars enrich the ISM at their death through \textit{(i)} radiative-driven mass loss; \textit{(ii)} core-collapse SNe; and -- after the explosion -- \textit{\textit{(iii)}} neutrino-driven winds from the proto-NS that may have been left after the explosion. To test this hypothesis, we also look at the abundances of Mo and Ru, which can be synthesized in p-rich outflows of proto-NSs \citep{hoffman1996,frohlich2006,pruet2006}. 
Throughout this paper, when we consider proto-NS winds we mean both purely neutrino-driven proto-NS winds and proto-magnetar models with strong magnetic fields and potentially rapid rotation as described in \citet{vlasov2017}. We emphasize the latter scenario, which achieves good agreement with
observations, while also showing that existing yield predictions for
the winds from unmagnetized, non-rotating proto-NSs overproduce some
observed abundances by a large factor \citep{hoffman1996,hoffman1997}. We note that other mechanisms mentioned above (e.g., NS-NS mergers, accretion disks, and magneto-rotational supernovae) may all play additional roles, which would require models for their frequencies and relative yields.
 Recent studies indicate that $\sim40$ per cent of NS births produce magnetar-strength fields \citep{beniamini2019}. Moreover, recent work by \citet{sukhbold2017} argues that normal Type IIP supernovae may arise from magnetars with few-ms spin period. Nevertheless, while the spin distribution of NSs with magnetar-strength fields is highly uncertain, the observed spin distribution of normal pulsars indicates that their NSs
are born rotating slowly; a typical average value of the initial spin period is $\langle P \rangle  = 300\,\text{ms}$, with $\sigma_{P} = 150\,\text{ms}$ \citep[table 6]{fausher2006}. 

Our paper is organized as follows. In Section \ref{sec:model} we describe our chemical evolution model. In Section \ref{sec:results} we present our results. Finally, in Section \ref{sec:conclusions}, we draw our conclusions. 

\section{The model} 
\label{sec:model}
We develop a chemical evolution model for the abundances in the ISM of our Galaxy, which makes the following assumptions for the various stellar and SN chemical enrichment sources. 

\begin{enumerate}

\item For AGB stars, we assume the stellar nucleosynthesis yields as computed by \citet{straniero2006},  \citet{cristallo2007,cristallo2009,cristallo2011,cristallo2015a,cristallo2015b,cristallo2016}, \citet{piersanti2013}, and \citet{straniero2014}, which are publicly available online at the FRUITY database website\footnote{\url{http://fruity.oa-teramo.inaf.it}}. Our assumed set of stellar yields for AGB stars includes the following metallicities $Z = 0.00002$, $0.00005$, $0.0001$, $0.0003$, $0.001$, $0.002$, $0.003$, $0.006$, $0.008$, $0.01$, $0.014$, $0.02$, with stellar masses in the range $1.3 < M < 6.0\;\mathrm{M}_{\sun}$. 
    
    \item We consider two alternatives for massive star yields. Our reference model assumes the stellar nucleosynthesis yields as computed by \citet{sukhbold2016} for non-rotating stars with masses in the range $9 \le M < 120\;\mathrm{M}_{\sun}$ at solar metallicity, which account for black hole formation and ``failed'' SNe (e.g., see \citealt{ugliano2012,pejcha2015,ertl2016,sukhbold2020} for simulations and models, and \citealt{smartt2009,gerke2015,adams2017,basinger2020} for an observational perspective). Specifically, we use the yields provided for the Z9.6 model below $12\,\text{M}_{\odot}$ and the W18 model at higher masses. \citet{sukhbold2016} compute yields at solar metallicity, which we apply at all metallicities. With the W18 central engine, most stars between $22$-$25\,\text{M}_{\odot}$ and above $28\,\text{M}_{\odot}$ form black holes without explosion, but they still produce enrichment through stellar winds. 
    
    Note that the nucleosynthesis calculations of \citet{sukhbold2016} do not account for neutrino interactions during the collapse and explosion, which can affect the evolution of $Y_{e}$ and hence the nucleosynthesis of the innermost ejecta, modifying the abundances of the immediate post-shock ejecta during the explosion (see \citealt{pruet2005,frohlich2005,frohlich2006,curtis2019,ebinger2020}), generating some p-rich elements. We do not include a complete description of the innermost ejecta, but we do adopt abundances for the p-rich elements Mo and Ru, which can be produced in p-rich proto-NS winds \citep{pruet2006}. In this way, our chemical evolution models combine the nucleosynthesis calculations of \citet{sukhbold2016} for core-collapse SNe with the models of \citet{pruet2006} for proto-NS winds, which include the potential for p-rich wind material. 
    
    Our alternative massive star yields use the calculations of \citet[set R]{limongi2018} with stellar rotation velocity $v_{\text{rot}} = 150\,\text{km}\,\text{s}^{-1}$ and the following grid of iron abundances: $[\text{Fe/H}] = -3$, $-2$, $-1$, $0$\footnote{These stellar yields are publicly available at the following website: \url{http://orfeo.iaps.inaf.it}}.  We note that \citet{limongi2018} also account for failed SNe, by assuming a sharp transition at $M > 25\,\text{M}_{\sun}$. Above this threshold mass, the chemical enrichment is provided only by the stellar winds.

    \item The average yields of Sr, Y, and Zr from neutrino-driven winds from proto-NSs are taken from the models of \citet{vlasov2017} with rotation periods $P=2$, $3$, $5$, and $10\,\text{ms}$. For Ba, our reference model assumes an average yield $M_{\text{Ba,r}} = 5\times10^{-7}\;\text{M}_{\sun}$, which is of the order of magnitude predicted in the winds of the most massive proto-NSs by \cite{wanajo2013}. Note that \citet{vlasov2017} find very little Ba production in all of their models. However, they did not systematically consider higher-mass proto-NSs and they did not include the effects of General Relativity (GR), both of which enhance heavy-element production \citep{cardall1997}, and were considered by \citet{wanajo2013}. Conversely, \citet{wanajo2013} did not consider the effects of a strong magnetic field and rapid rotation on the nucleosynthesis, which \citet{vlasov2017} find have important consequences for production of elements $Z<55$. For these reasons, we take the Sr, Y, Zr abundances from \citet{vlasov2017} and take the Ba abundances from \citet{wanajo2013} under the assumption that \citet{vlasov2017}-like models including GR and a range of higher proto-NS masses would yield results for Ba more like the \citet{wanajo2013} yields. 
    
    The calculations of \citet{wanajo2013} adopt a fixed $Y_{e}=0.4$ that is optimistic from the point of view of $r$-process nucleosynthesis. The value $Y_e=0.4$ is near, but below, the most neutron-rich value of $Y_e$ obtained in the proto-NS cooling calculations used in \citet{vlasov2017}. It is not clear whether such low values of $Y_e$ can be attained generically for all proto-NS masses and rotation rates. Conversely, the models of \citet{vlasov2017} adopt the full proto-NS cooling calculation of \citet{roberts2010} for a non-rotating $1.4\,\text{M}_{\sun}$ proto-NS, which represents an up-to-date and modern calculation predicting low $Y_{e} < 0.5$ in the wind cooling epoch. However, the predicted $Y_{e}$ never decreases below $0.44$ and eventually becomes proton-rich after approximately $6$\,seconds of cooling. It is unclear how the proto-NS cooling calculations would generalize to the range of masses considered by \citet{wanajo2013} or the rapid rotation and strong magnetic effects employed by \citet{vlasov2017}, since one expects the details of the cooling process, and the resulting wind $Y_e$ to depend on both mass and rotation. Because \citet{wanajo2013} uses a fixed, optimistic value of $Y_e=0.4$ for a range of masses, we use it as a measure of the uncertainty in the Ba production. The details of both the wind and explosion nucleosynthesis can be further modified by multi-dimensional effects such as continued accretion during the explosion as the wind emerges; in this regard, \citet{bollig2021} suggest that the wind develops only in the lowest-mass progenitor cases due to a long-term accretion flow, which hinders the neutrino-driven wind. All these issues point to the urgent need for a next generation of magnetic and rapidly rotating  proto-NS wind models including GR, a range of proto-NS masses, and dynamical magnetospheres (see \citealt{thompson2018}).
    
   Since the wind calculations of \citet{vlasov2017} do not address the physics of the $\nu$p-process giving rise to high-$Y_{e}$ early ejecta, we adopt the models of \citet{pruet2006} for the chemical elements Mo and Ru. In particular, the production factors for the most abundant isotopes of Mo and Ru, $P_{\text{Mo,Ru}}$, are in the range $10$-$30$, according to the predictions of the models of \citet{pruet2006}, with the average yield of $i =$ Mo, Ru being defined as \begin{equation} \label{eq:equation1}
         M_{i,\text{pnw}}(m) = P_{i} \times X_{\sun,i} \times M_{\text{tot-ej,SN}}(m), 
    \end{equation} where $X_{\sun,i}$ is the Solar abundance by mass of $i =$ Mo, Ru \citep{asplund2009}, and $M_{\text{tot-ej,SN}}(m)$ is the total ejected mass in the core-collapse SN explosion of a massive star with initial mass $m$.  
    
    \item For Type Ia SNe, we assume the empirically motivated delay-time distribution function (DTD) \citep{maoz2012}: \begin{equation}
        \text{DTD}_{\text{Ia}}(t) = A_{\text{Ia}} \, t^{-1.1},
    \end{equation} where $A_{\text{Ia}}$ is chosen in order to have two Type Ia SNe over $13.8\,\text{Gyr}$ per $10^{3}\,\text{M}_{\sun}$ of stellar mass formed \citep{bell2003,maoz2014,vincenzo2017}. In this work, we assume a minimum delay-time $\tau_{\text{min,Ia}} = 150\;\text{Myr}$. The stellar yields of Type Ia SNe are from \citet{iwamoto1999}. 

\end{enumerate}

\textit{The star formation rate} --- We assume that the star formation rate (SFR) follows a linear Kennicutt law, namely $\text{SFR}(t) = \text{SFE}\times M_{\text{gas}}(t)$, where $\text{SFE}$ represents the star-formation efficiency and $M_{\text{gas}}(t)$ is the total gas mass in the Galaxy. 

\textit{The accretion rate} --- The simulated galaxy is assumed to form from the accretion of primordial gas from the circumgalactic environment. The gas infall rate obeys the following law: $\mathcal{I}(t) = A_{\mathcal{I}} \times e^{-t/\tau_{\text{inf}}}$, where $A_{\mathcal{I}}$ is a normalization constant that determines the total amount of gas accreted over the Hubble time, defining the so-called infall-mass, $M_{\text{inf}}$, of the models. 

\textit{The outflow rate} --- The model can account for the effect of galactic outflows, which carry gas and metals out of the galaxy potential well. The galactic winds are assumed to proceed with a rate proportional to the SFR, namely $\mathcal{O}(t) = \omega \text{SFR}(t)$, where $\omega$ is the so-called mass-loading factor. We note in advance that our reference chemical evolution model -- for simplicity -- does not assume outflow activity ($\omega = 0$).  

\textit{The chemical-enrichment rate} --- The chemical enrichment from winds of AGB stars and massive stars, core-collapse SNe, Type Ia SNe, and neutrino-driven winds from proto-NSs is included in our chemical evolution model according to the following equation:
\begin{equation} \label{eq:chem-ev}
\begin{aligned}
    \mathcal{R}_{i}(t) &= \int\displaylimits_{m_{\text{TO}}(t)}^{m_{\text{cutoff}}}{ \mathrm{d}m \, \text{IMF}(m) \, \text{SFR}(t - \tau_{m}) \, p_{i}\big( m , Z( t - \tau_{m} ) \big) } \\ 
    &+ p_{i,\text{Ia}} \, \int_{\tau_{\text{min,Ia}}}^{t} \text{d}\tau \, \text{DTD}_{\text{Ia}}(\tau)\,\text{SFR}(t-\tau), 
\end{aligned}
\end{equation}
which describes the chemical-enrichment rate of the $i$-th chemical element at the time $t$ from all the assumed dying nucleosynthetic sources. In the reference model, we assume the initial mass function (IMF) of \citet{kroupa1993} and the stellar lifetimes, $\tau_{m}$, of \citet{kobayashi2004}. Note that the \citet{kroupa1993} IMF is quite different from the one usually referred to as a ``Kroupa IMF'' \citep{kroupa2001}. It has a high mass slope of $-2.7$ rather than $-2.3$, and it therefore predicts lower IMF-averaged yields from massive stars for a given SFR.

The quantity $m_{\text{TO}}(t)$ in equation (\ref{eq:chem-ev}) represents the turn-off mass, which is derived from the assumed inverse stellar lifetimes of \citet{kobayashi2004}; the quantity $m_{\text{cutoff}} = 100\,\text{M}_{\sun}$ represents the maximum stellar mass, which is assumed to form in the star-formation events; the quantity $p_{i}(m,Z)$ represents the stellar nucleosynthetic yields of the $i$-th chemical element from all stars with mass $m$ and metallicity $Z$; finally, $p_{i,\text{Ia}}$ is the average nucleosynthesis yield from Type Ia SNe. Note that $p_{i}(m,Z)$ includes the nucleosynthesis yields from stellar winds of AGB stars and massive stars, core-collapse SNe, and neutrino-driven winds from proto-NSs. We assume that proto-NS winds arise for stars with progenitor masses $8 \le m <20\;\mathrm{M}_{\sun}$ and that more massive progenitors produce black hole remnants with no proto-NS winds.

\textit{The chemical evolution equations} --- Using all quantities defined above, our model solves the following differential equation for the evolution of the gas mass in the form of the $i$-th chemical element in the ISM of the galaxy as a function of time:
\begin{equation}
    \frac{ \text{d}M_{\text{g},i}(t) }{\text{d}t} = -X_{i}(t)\,\text{SFR}(t) + \mathcal{R}_{i}(t) + \mathcal{I}(t) - \mathcal{O}(t), 
\end{equation}
where $X_{i}(t)= M_{\text{g},i}(t)/M_{\text{gas}}(t)$ is the ISM abundance by mass of the $i$-th chemical element at the time $t$. In our reference chemical evolution model, we assume $\text{SFE}=2\,\text{Gyr}^{-1}$, $\tau_{\text{inf}}=5\;\text{Gyr}$, $\omega = 0$,  $\log(M_{\text{inf}}/\text{M}_{\sun})= 11.5$, which are tuned to reproduce the observed chemical abundance patterns of [$X$/Fe]-[Fe/H], where $X=$ O, Sr, Y, Zr, Ba, Mo, and Ru, at the Solar neighbourhood. These values are consistent with previous works (e.g., \citealt{minchev2013,nidever2014,spitoni2015,vincenzo2017,magrini2018}). In contrast to the models of \citet{andrews2017} and \citet{weinberg2017}, we are able to obtain solar metallicities with $\omega=0$ because the steeper IMF and reduction of massive star yields by black hole formation lowers the IMF-averaged yields by a substantial factor. We run the model for a time interval of $10\,\text{Gyr}$.

\begin{figure}
\centering
\includegraphics[width=8.0cm]{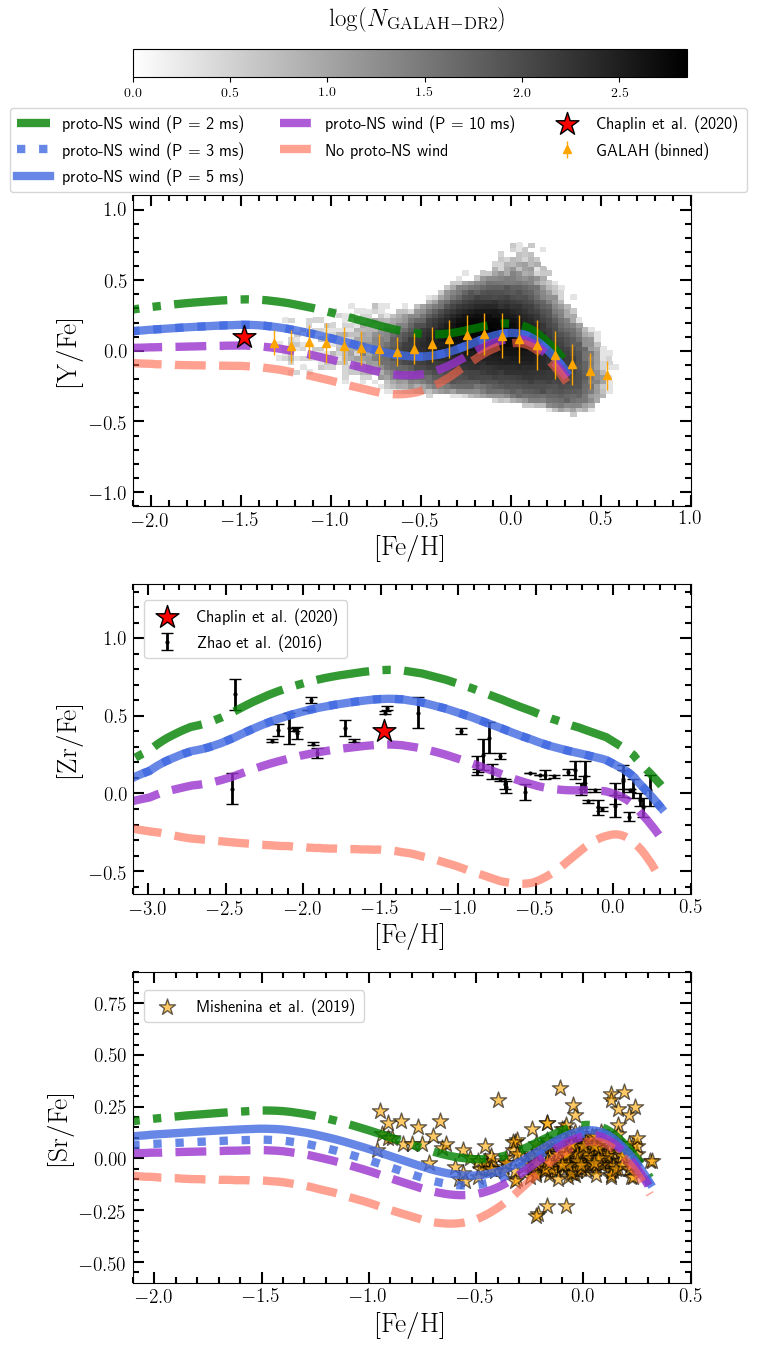}
\caption{The predictions of chemical evolution models including a contribution to Sr, Y, and Zr from neutrino-driven winds of proto-NSs with different rotation period \citep{vlasov2017} (green dashed-dotted line: $P=2\,\text{ms}$; blue dotted line: $P=3\,\text{ms}$; blue solid line: $P=5\,\text{ms}$; magenta solid line: $P=10\,\text{ms}$). These models are compared with a model which does not assume proto-NS winds (red dashed curve), but only chemical enrichment from stellar winds of AGB stars \citep{cristallo2016} and massive stars \citep{sukhbold2016}, and core-collapse SNe \citep{sukhbold2016}. The observational data are from \citet[black error bars]{zhao2016}, \citet[orange star symbols]{mishenina2019}, and \citet[red star symbol]{chaplin2020}. For Y, we also show the observational data from GALAH-DR2 \citep{buder2018}, which are represented by the gray colour-code two-dimensional histogram, as well as by the average binned data (yellow triangles with error bars), by selecting only dwarf stars with $\log(g) > 3.5$ and signal-to-noise ratio $\text{SNR}>20$. Model curves (not shown) that adopt the predicted yields of spherical winds from unmagnetized proto-NSs would be mostly off the top of these plots, exceeding the observed abundance ratios by $1$-$1.5$ dex at all metallicities.   }
\label{fig:first-peak}
\end{figure}

\begin{figure}
\centering
\includegraphics[width=8.0cm]{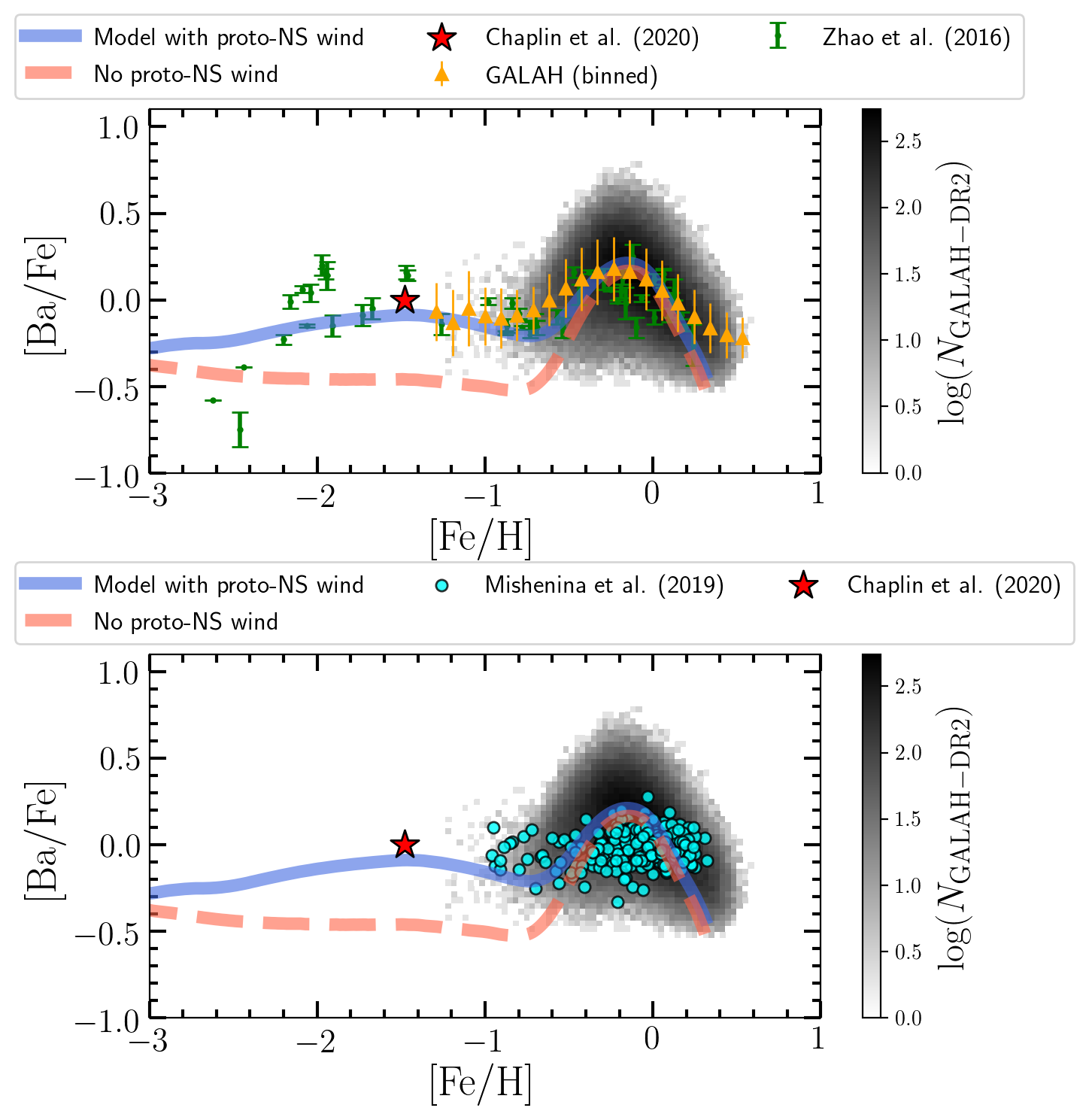}
\caption{The predicted [Ba/Fe]-[Fe/H] abundance pattern from our reference chemical evolution model including neutrino-driven winds of massive proto-NSs (blue solid curve) as compared with a model which does not assume proto-NS winds (red dashed curve, like in Fig. \ref{fig:first-peak}). For these predictions we base the Ba yield on the calculations of 
\citet{wanajo2013} rather than the much lower Ba yields computed by \citet{vlasov2017} (see \S 2). The two panels focus on different sets of observational data, which are from \citet[green error bars]{zhao2016}, \citet[cyan filled circles]{mishenina2019}, and \citet[red star symbol]{chaplin2020}. The observational data from GALAH-DR2 \citep{buder2018} are shown as in Fig. \ref{fig:first-peak}.   }
\label{fig:ba}
\end{figure}

\begin{figure}
\centering
\includegraphics[width=8.0cm]{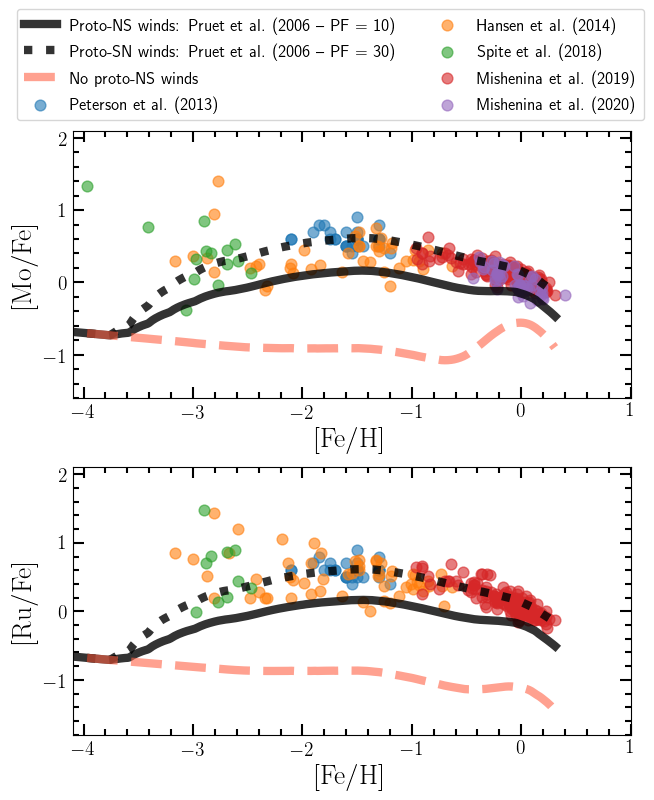}
\caption{Predictions for [Mo/Fe]-[Fe/H] (upper panel) and [Ru/Fe]-[Fe/H] (lower panel). The black curve correspond to the models with p-rich outflows from proto-NSs \citep{pruet2006}, by assuming a production factor (PF) of 10 (black solid curve) and $30$ (black dashed curve). The model without the contribution of proto-NS winds is the red dashed line, like in Fig. \ref{fig:first-peak}. The observational data are from \citet[blue filled circles]{peterson2013}, \citet[orange filled circles]{hansen2014}, \citet[green filled circles]{spite2018}, \citet[red filled circles]{mishenina2019}, \citet[violet filled circles]{mishenina2020} }
\label{fig:mo-ru}
\end{figure}

\begin{figure}
\centering
\includegraphics[width=8.5cm]{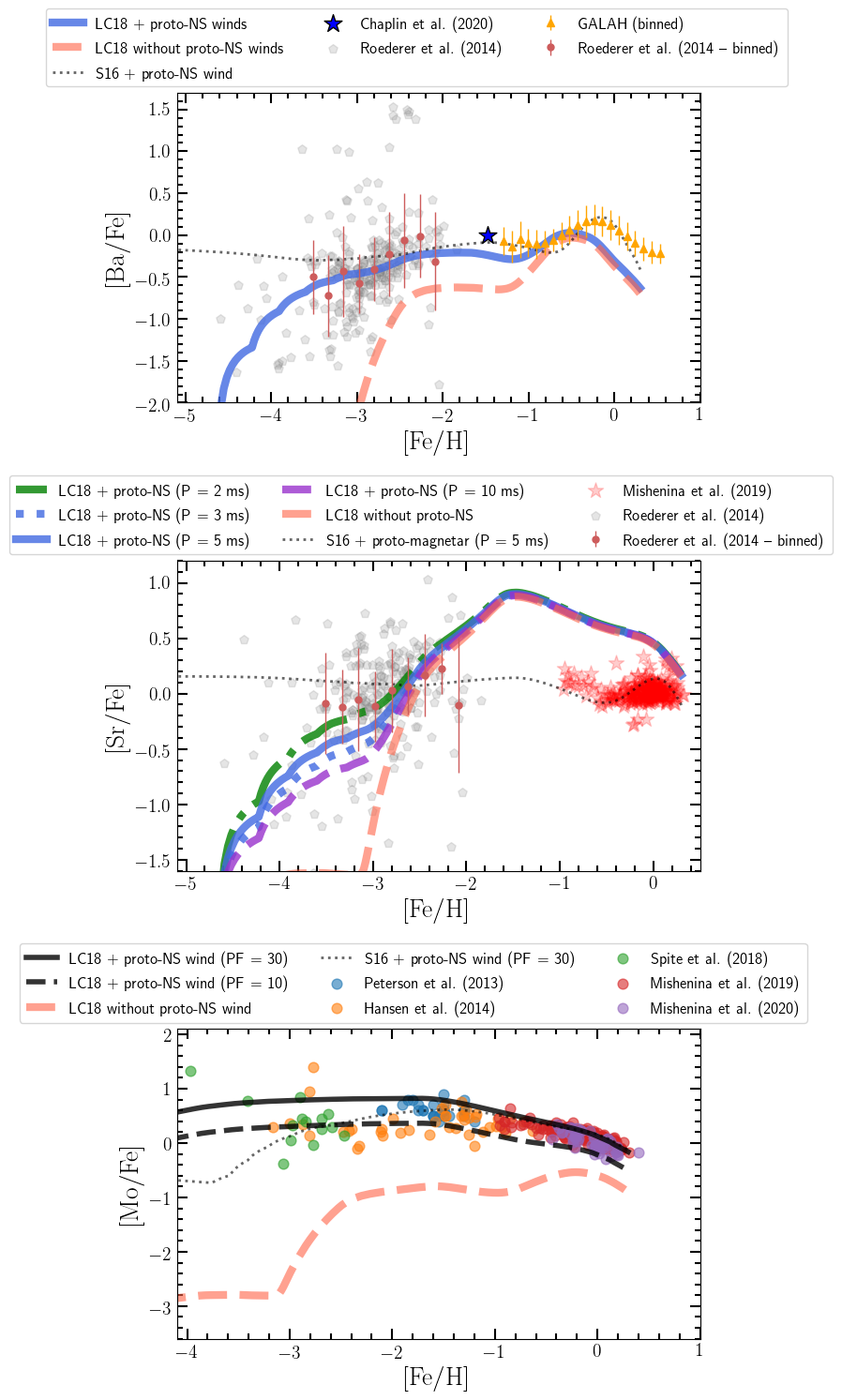}
\caption{Predictions for [Ba/Fe]-[Fe/H] (upper panel), [Sr/Fe]-[Fe/H] (middle panel), and [Mo/Fe]-[Fe/H] (lower panel), focusing on the trend at low metallicities in order to reproduce the observations of \citet[gray pengatons]{roederer2014}, which have also been binned in the range $-3.6\le\text{[Fe/H]}<-2.0$ to determine the mean trend and the dispersion of the data (red point with error bars). We show models assuming the stellar yields of \citet[LC18; set R]{limongi2018} for massive stars ($v_{\text{rot}}=150\,\text{km}\,\text{s}^{-1}$), the AGB stellar yields of \citet{cristallo2016}, and neutrino-driven winds from proto-NSs (Sr from \citealt{vlasov2017}, Ba from \citealt{wanajo2013}, and Mo from \citealt{pruet2006}). The various curves and the remaining observational data are the same as in Fig. \ref{fig:ba}. We also show for comparison the prediction of the model with \citet[S16; thin black dotted curve]{sukhbold2016} yields with chemical enrichment from massive stars and proto-NS winds with $P=5\,\text{ms}$. The \citet{sukhbold2016} yields are computed only for solar metallicity.}
\label{fig:ba-sr-v150}
\end{figure}

\begin{figure}
\centering
\includegraphics[width=8.0cm]{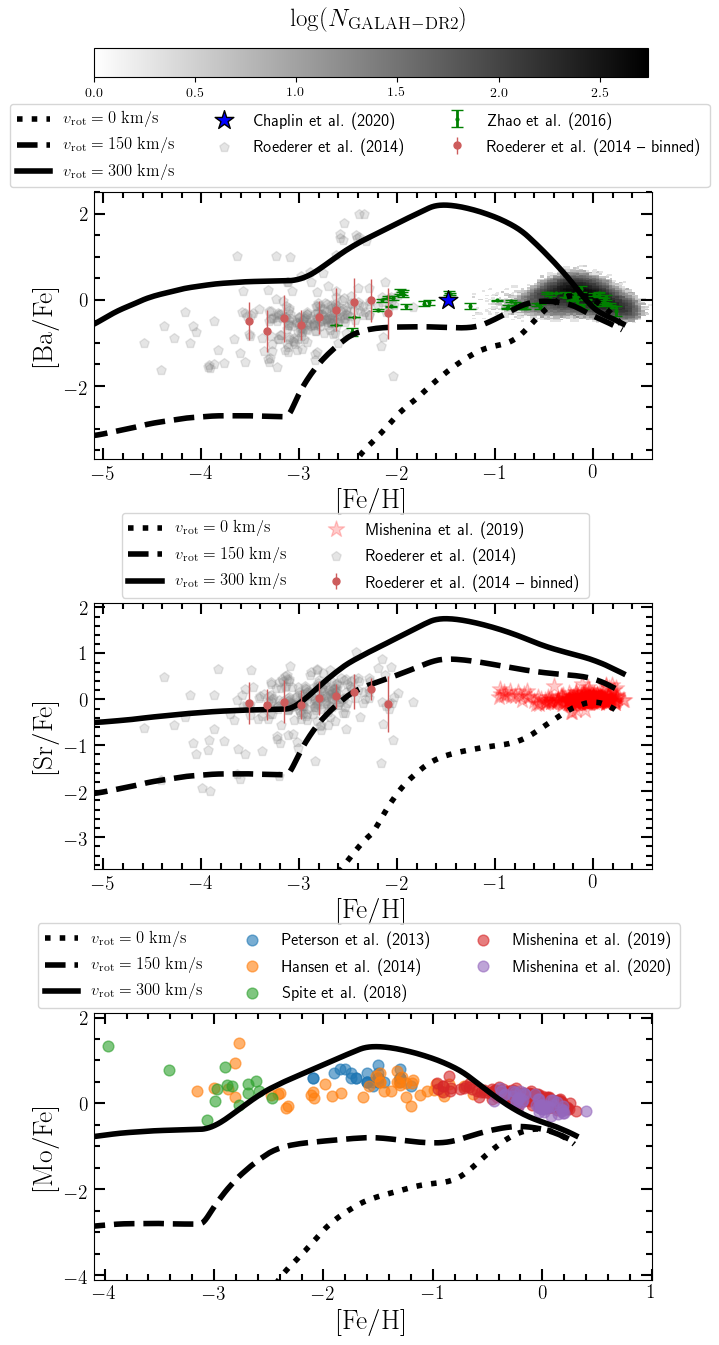}
\caption{Model predictions of [Ba/Fe] (upper panel), [Sr/Fe] (intermediate panel), and [Mo/Fe] (bottom panel) as a function of [Fe/H], as computed by using the nucleosynthetic stellar yields of \citet{limongi2018} models. The black dotted curve corresponds to $v_{\text{rot}}=0\,\text{km}\;\text{s}^{-1}$, the black dashed curve to $v_{\text{rot}}=150\,\text{km}\;\text{s}^{-1}$, and the black solid to $v_{\text{rot}}=300\,\text{km}\;\text{s}^{-1}$. The observational data are the same as in Figs. \ref{fig:ba}-\ref{fig:ba-sr-v150}. }
\label{fig:ba-sr-IMFaveraged}
\end{figure}

\section{Results} \label{sec:results}
The predictions of our reference chemical evolution model for the first peak s-process elemental abundance ratios are shown in Fig. \ref{fig:first-peak}. For [Y/Fe], the observational data are from the second data release of the GALactic Archaeology with HERMES (GALAH) spectroscopic survey \citep[two dimensional histogram with the gray color-coding]{buder2018}, considering only dwarf stars with surface gravity $\log(g) > 3.5$ and signal-to-noise ratio $\text{SNR}>20$. To understand the average trend of the data, we also show the mean [Y/Fe] from GALAH with the corresponding $\pm 1\sigma$ dispersion as a function of [Fe/H] (yellow triangles with error bars). Finally, we show the [Y/Fe] ratio as measured by \citet{chaplin2020} in a bright star belonging to the inner MW halo ($\nu$ Indi) (red star symbol), for which they measured an asteroseismic age of $\approx11\;\text{Gyr}$ from the analysis of the TESS oscillation spectrum of the star. For [Zr/Fe], the observational data are from \citet[red star symbol]{chaplin2020} and \citet[black data with error bars]{zhao2016} for a sample of stars in the Solar neighbourhood. For [Sr/Fe] we show the abundance measurements of \citet[yellow star symbols]{mishenina2019}. All observational data shown in Fig. \ref{fig:first-peak} are measured by accounting for non-local thermodynamic equilibrium (NLTE) effects in the abundance analysis. 

When accounting only for the chemical enrichment from core-collapse SNe, Type Ia SNe, and the stellar winds of AGB stars and non-rotating massive stars (pink dashed curve in Fig. \ref{fig:first-peak}), we cannot reproduce the observed abundance ratios of [X/Fe] of the first-peak s-process elements ($X=$ Sr, Y, Zr). In particular, our chemical evolution model consistently underpredicts the abundance ratios at metallicities below $\text{[Fe/H]}\approx-0.5$. Including the additional r-process contribution from neutrino-driven winds of proto-NSs with rotation periods in the range $2 \le P \le 5\,\text{ms}$, significantly improves the agreement with the data. For $P=10$ ms the model underpredicts the observed [Sr/Fe] ratios
at $\text{[Fe/H]} \approx -1$, though agreement with [Y/Fe] and [Zr/Fe] is
acceptable.  For $P=10$ ms the proto-NS is effectively ``non-rotating''
in the sense that the predicted yields would not decrease much
for still longer periods.  

Although we do not show them in Fig.~\ref{fig:first-peak}, we have also computed models using
the yields of the ``spherical'' calculations of \cite{vlasov2017} for
non-rotating, unmagnetized proto-NS.  These models overpredict the
observed [Y/Fe], [Zr/Fe], and [Sr/Fe] ratios by $1$-$1.5$ dex over the
entire metallicity range shown in Fig.~\ref{fig:first-peak}, so they are very clearly
ruled out, as anticipated by earlier studies of neutrino-driven winds
\citep{woosley1994,hoffman1996,hoffman1997,roberts2010}.  As discussed by \cite{vlasov2017}, magnetic fields sharply reduce the mass ejected by proto-NS winds because of the small fraction of the stellar surface
threaded by open magnetic flux lines.  
The conflict between spherical model yields and observed
abundances suggests that most proto-NSs are indeed significantly magnetized.
Alternatively, some currently unrecognized neutrino transport physics
could alter the ratio of electron- and anti-electron neutrino fluxes
in the early cooling epoch of the proto-NS wind ($t \la 2\,$s),
changing the predicted nucleosynthesis by altering the electron fraction
along a given thermodynamic trajectory of the expanding matter.

In Fig. \ref{fig:first-peak}, the predicted flat trend of [Y/Fe] without proto-NS winds (red dashed curve) at $[\text{Fe/H}] \lesssim -2.0$ is due to the chemical enrichment of Y from the winds of massive stars and Fe from core-collapse SNe. At metallicities in the range $-1.2 \lesssim [\text{Fe/H}] \lesssim -0.6$, [Y/Fe] decreases because of the large amounts of Fe injected in the ISM by Type Ia SNe. At $[\text{Fe/H}]\approx - 0.6$, [Y/Fe] increases because of the large amounts of Y produced by AGB stars per unit time, which is eventually overcome by Type Ia SNe at $[\text{Fe/H}]\gtsim 0$. A similar explanation is valid for the predicted trends of [Sr/Fe] and [Zr/Fe], in agreement with discussion of [Sr/Fe] by \citet{johnson2020}. 
However, without a proto-NS wind contribution, these trends lie below the data by $\sim0.3$-$1\,\text{dex}$.

In Fig. \ref{fig:ba} we compare the predictions of our reference chemical evolution model for [Ba/Fe] with the observational data. Models with an additional r-process contribution from neutrino-driven winds of massive proto-NSs ($M_{\text{Ba,r}}=5\times10^{-7}\,\text{M}_{\sun}$ per event) provide excellent agreement with the bulk of the observational data from the GALAH survey \citep{buder2018} as well as with \citet{zhao2016} and \citet{chaplin2020}, all including NLTE effects in their abundance analysis. Without this contribution, the model underpredicts observed abundances at $\text{[Fe/H]}\lesssim-1$. 

It has been shown that the p-rich outflows of proto-NSs can be highly effective in the nucleosynthesis of Mo and Ru \citep{pruet2006}. In order to test this scenario, in Fig. \ref{fig:mo-ru} we compare the predictions of our models for [Mo/Fe] (upper panel) and [Ru/Fe] (lower panel) with a set of observational data  \citep{peterson2013,hansen2014,spite2018,mishenina2019,mishenina2020}. Similarly to what we find for the first- and second-peak s-process elements, our chemical evolution model including proto-NS winds with a production factor $P=30$ produces a good match to the bulk of the observational data of [Mo/Fe] and [Ru/Fe] as a function of [Fe/H], while the model without proto-NS winds falls far short.

 \begin{table} \label{table1}
\centering  
\begin{tabular}{c|c|c|c|c|c}
\hline
\hline
\multicolumn{6}{c}{$\mathbf{\text{\bf[Fe/H]} = -3.0}$} \\
 $v_{\text{rot}}/\text{[km/s]}$  & [Sr/Fe] & [Y/Fe] & [Zr/Fe] & [Ba/Fe] & [Mo/Fe] \\
\hline

$0$  & $-3.55$ & $-3.58$ & $-3.59$ & $-3.59$  & $-3.55$ \\

$150$  & $-1.97$ & $-2.22$ & $-2.35$ & $-2.85$ & $-2.79$  \\

$300$  & $1.10$ & $-0.45$ & $-0.24$ & $0.10$ & $-0.21$  \\

\hline
\hline

\multicolumn{6}{c}{$\mathbf{\text{\bf[Fe/H]} = -2.0}$} \\
 $v_{\text{rot}}/\text{[km/s]}$  & [Sr/Fe] & [Y/Fe] & [Zr/Fe] & [Ba/Fe] & [Mo/Fe] \\
\hline

$0$    & $-2.49$ & $-2.54$ & $-2.58$ & $-2.57$ & $-2.56$  \\

$150$  & $-0.14$ & $-0.25$ & $-0.42$ & $-0.98$ & $-1.00$  \\

$300$  & $1.41$ & $0.57$ & $0.85$ & $1.10$ & $0.62$  \\

\hline
\hline

\multicolumn{6}{c}{$\mathbf{\text{\bf[Fe/H]} = -1.0}$} \\
 $v_{\text{rot}}/\text{[km/s]}$  & [Sr/Fe] & [Y/Fe] & [Zr/Fe] & [Ba/Fe] & [Mo/Fe] \\
\hline

$0$    & $-1.07$ & $-1.23$ & $-1.38$ & $-1.48$ & $-1.48$  \\

$150$  & $0.58$ & $0.33$ & $0.01$ & $-0.87$ & $-0.74$  \\

$300$  & $0.41$ & $1.51$ & $1.81$ & $2.02$ & $1.51$  \\

\hline
\hline

\multicolumn{6}{c}{$\mathbf{\text{\bf[Fe/H]} = 0}$} \\
 $v_{\text{rot}}/\text{[km/s]}$  & [Sr/Fe] & [Y/Fe] & [Zr/Fe] & [Ba/Fe] & [Mo/Fe] \\
\hline

$0$    & $-0.35$ & $-0.44$ & $-0.51$ & $-0.54$ & $-0.54$  \\ 

$150$  & $0.37$ & $0.20$ & $-0.05$ & $-0.45$ & $-0.61$  \\

$300$  & $0.39$ & $0.71$ & $0.51$ & $-0.20$ & $-0.32$  \\

\hline
\hline

\end{tabular}
\caption{ The IMF-averaged yield of [X/Fe] ($X=$ Sr, Y, Zr, Ba, and Mo) as predicted by the massive star models of \citet{limongi2018}, as a function of [Fe/H] and rotation velocity. The assumed IMF is that of \citet{kroupa1993}. The values in the table are computed by using \textit{gross} stellar nucleosynthetic yields, namely including also the contribution of the ejected material which was present at the stellar birth and remained unprocessed. The IMF-averaged stellar yields are computed by using the code {\tt VICE} \citep{johnson2020}. These yields are reported in terms of [X/Fe], where Fe is the IMF-averaged iron yield from massive stars only.
}
\end{table}

\subsection{The impact of rotating massive stars at low metallicities}

The \citet{sukhbold2016} yields are available only at solar metallicity, and they assume non-rotating massive star progenitors. To address both of these potential shortcomings, we consider the alternative yields of \citet{limongi2018}. Table 1 reports the IMF-averaged [X/Fe] ratios ($X=$ Sr, Y, Zr, Ba, and Mo) as predicted when assuming these yields, for different [Fe/H] abundances and rotation velocities.
Larger rotation velocity causes an enhancement of the s-process production in massive stars (see also figure 4 of \citealt{johnson2020}); this is due to the so-called rotation-induced mixing, which can bring material from the convective H-burning shell (in particular, $^{14}$N nuclei produced in the CNO cycle) to the He-burning core, where the reaction $^{22}$Ne($\alpha$, $n$)$^{25}$Mg takes place (the main source of free neutrons in massive stars). When the $^{14}$N nuclei reach the He-core, they can capture two $\alpha$-particles to produce $^{22}$Ne, eventually giving rise to more s-process events. This physical mechanism, which enhances the s-process nucleosynthesis in massive stars, was originally proposed by \citet{frischknecht2012} to explain the s-process nucleosynthesis at very low-metallicity (see also \citealt{cescutti2013}). 

The rotation velocity of massive stars is highly uncertain. One benchmark study is that of \citet{agudelo2013}, who found that the distribution of the projected rotation velocities in a sample of massive stars in the Tarantula Nebula has a peak at $\sim80\,\text{km/s}$, with the $80$th percentile being at $\approx300\,\text{km/s}$. Therefore, in the context of the stellar models of \citet{limongi2018}, a value of $v_{\text{rot}}\approx300\;\text{km/s}$ should be considered as an approximate upper limit from an observational point of view, with the majority of the stars likely rotating with velocities in the range $0 < v_{\text{rot}}<150\;\text{km/s}$. However, typical rotation speeds could be different at very low metallicities. 

In Fig. \ref{fig:ba-sr-v150}, we show the predictions of chemical evolution models assuming the stellar nucleosynthesis yields of \citet{limongi2018} for $v_{\text{rot}}=150\,\text{km/s}$ with and without proto-NS winds (blue solid curve and red dashed curve, respectively, for [Ba/Fe] in the upper panel). We assume yields are constant below $\text{[Fe/H]}=-3$. For comparison, we also show our reference chemical evolution model assuming the stellar yields of \citet{sukhbold2016} with proto-NS winds ($P=5\,\text{ms}$; thin dotted line), taking these yields to be metallicity independent. 
If we do not include proto-NS winds, the predicted [Sr/Fe], [Ba/Fe], and [Mo/Fe] lie well below the \citet{roederer2014} data at $-4 \le[\text{Fe/H}] \le -2$. Conversely, when we include our standard estimate of the r-process contribution from proto-NS winds, we obtain better agreement with the average trend of the observational data at low [Fe/H]. The [Sr/Fe] comparison prefers the higher proto-NS yields of the $P=2\,\text{ms}$ models. The chemical evolution model assuming rotating massive stars with $v_{\text{rot}}=150\,\text{km/s}$ and proto-NS winds systematically overestimates [Sr/Fe] at $[\text{Fe/H}] \gtrsim -2$. This disagreement can be alleviated by transitioning to low rotation speeds for $[\text{Fe/H}]\gtrsim-2$, thus moving towards the black dashed curve computed with \citet{sukhbold2016} stellar yields. 

If we assume still higher rotation velocities at low metallicity, then it becomes possible to reproduce the data without the addition of proto-NS winds. Fig. \ref{fig:ba-sr-IMFaveraged} compares the observed chemical abundance ratios of [Ba/Fe] (upper panel), [Sr/Fe] (intermediate panel), and [Mo/Fe] (bottom panel) with the predictions of models assuming the stellar nucleosynthetic yields of \citet{limongi2018} for $v_{\text{rot}}=0$, $150$, and $300\,\text{km}\,\text{s}^{-1}$. At least in overall level, the observed [Sr/Fe] and [Ba/Fe] ratios at low metallicities can be explained by models with rotation velocities in the range $150 < v_{\text{rot}} < 300\;\text{km/s}$, without the need of a significant r-process contribution from additional sources. 
Reproducing the observed [Mo/Fe] requires typical rotation speeds at the top of this range.

\section{Conclusions} \label{sec:conclusions}

In this work, we have tested the hypothesis that the observed abundances of first-peak (i.e. Sr, Y, Zr) and second-peak (i.e. Ba) s-process elements, as well as the abundances of Mo and Ru, can be explained by incorporating an additional r-process contribution at low metallicities from neutrino-driven winds from proto-NSs. To this aim, we have developed chemical evolution models for the evolution of the elemental abundances in our Galaxy including proto-NS winds, also investigating the impact of different assumptions for the chemical enrichment of massive stars, which can be important s-process contributors of light neutron-capture elements at low [Fe/H] \citep{frischknecht2012,cescutti2013}. 

We base our proto-NS wind yields on the calculations of \citet[for Y, Sr, Zr]{vlasov2017}, \citet[for Ba]{wanajo2013}, and \citet[for Mo, Ru]{pruet2006}. We caution that the \citet{wanajo2013} Ba yield of $\approx 5\times 10^{-7} M_\odot$ per event is much higher than the Ba yield predicted by \cite{vlasov2017}, so we regard our Ba predictions as more uncertain (see Section \ref{sec:model} for details). For massive stars, we construct models using the non-rotating models of \citet{sukhbold2016} computed at $\text{[Fe/H]}=0$, and alternative models using the yields of \citet[set R]{limongi2018}, which are available for different rotation velocities and [Fe/H] abundances. Both \citet{sukhbold2016} and \citet{limongi2018} massive star models account for failed SNe, but \citet{sukhbold2016} compute an explosion landscape based on a neutrino-driven central engine, while \citet{limongi2018} impose a mass threshold for black hole formation at $25\,\text{M}_{\sun}$.

Our main conclusion can be summarized as follows.
\begin{enumerate}

\item Adding the predicted proto-NS wind yields to the \cite{sukhbold2016}
massive star yields, and our standard choice of Type Ia supernova and AGB
yields, leads to good agreement with the observed trends of [Y/Fe], [Zr/Fe],
[Sr/Fe], [Ba/Fe], [Mo/Fe], and [Ru/Fe] (see Figs.~\ref{fig:first-peak}-\ref{fig:mo-ru}). 
The best agreement for Sr, Y, and Zr is obtained
for proto-NS rotation periods $P \sim 2-5$ ms, while models with
$P \sim 10$ ms (which are effectively in the non-rotating
limit for our purposes) underpredict the observed [Sr/Fe]. For Mo and Ru,
production factors of 10-30 (see equation \ref{eq:equation1}) are required, similar
to the proto-NS wind predictions of \cite{pruet2006}. Without  proto-NS
winds the models underpredict the observations by $0.3-1$ dex for stars
with $\text{[Fe/H]} \la -0.5$, though for Y, Sr, and Ba the AGB contribution
leads to acceptable agreement near solar metallicity.  

\item Because the \cite{sukhbold2016} yields assume non-rotating, solar
metallicity progenitors, we have also considered the alternative yield
sets of \cite{limongi2018} for $\text{[Fe/H]} = -3,-2,-1,0$ and progenitor rotation
velocities of $v_{\text{rot}} = 0, 150, 300\,\text{km}\,\text{s}^{-1}$.  For $v_{\text{rot}}=150\,\text{km}\,\text{s}^{-1}$, we find
reasonable agreement with observed [Ba/Fe] and [Sr/Fe] trends for
$-4 < \text{[Fe/H]} < -2$ with the addition of proto-NS wind yields with
$P=2-5$ ms (see Fig. \ref{fig:ba-sr-v150}).  Without proto-NS winds, these models severely underpredict
the observed Ba and Sr abundances at $\text{[Fe/H]} < -2$.

\item In the range $-1 \la \text{[Fe/H]} \la -0.5$, models with the \cite{limongi2018},
$v_{\text{rot}} =150\,\text{km}\,\text{s}^{-1}$ overpredict the observed [Sr/Fe], even without proto-NS
winds (see Fig. \ref{fig:ba-sr-v150}).  This conflict suggests that rotation velocities of massive stars
must be lower than $150\,\text{km}\,\text{s}^{-1}$ at these metallicities.

\item The predicted s-process yields of low metallicity massive stars are
sensitive to rotation, increasing by 1-3 orders of magnitude for
$v_{\text{rot}}=300\,\text{km}\,\text{s}^{-1}$ vs.\ $v_{\text{rot}}=150\,\text{km}\,\text{s}^{-1}$ (see Table 1).  Even without proto-NS winds,
models with yields intermediate between these two cases could reproduce
the observed levels of Ba and Sr at $\text{[Fe/H]} < -2$ (see Fig. \ref{fig:ba-sr-IMFaveraged}).  A model with
$v_{\text{rot}}=300\,\text{km}\,\text{s}^{-1}$ yields could reproduce the observed levels of Mo,
though not the detailed trend.  The observations of \cite{agudelo2013}
in the Tarantula nebula favor typical rotation velocities $< 150\,\text{km}\,\text{s}^{-1}$,
but higher rotation might be possible at low metallicity because of
reduced mass loss and associated angular momentum loss.  Very high
rotation velocities are disfavoured in some recent chemical evolution
models such as those of \cite{prantzos2018,prantzos2020} and
\cite{kobayashi2020}.

\item Models that adopt the \cite{vlasov2017} yields for spherical
winds from non-rotating, unmagnetized proto-NS are strongly ruled
out, overpredicting the observed Sr, Y, and Zr abundances by 1-1.5 dex,
in agreement with previous studies \citep{woosley1994,hoffman1996,hoffman1997,roberts2010}. 

\end{enumerate}

In summary, the winds from proto-NS with rotation periods $P\sim 2-5$ ms
offer a natural explanation for the observed abundances of Sr, Y, Zr, Mo, Ru, and
potentially Ba in Milky Way stars with $\text{[Fe/H]} < -0.5$, where models without
this contribution fall short by $0.3$-$1\,\text{dex}$ or even more.  Models with
rapidly rotating massive stars might be able to reproduce the observed
trends without the addition of proto-NS winds, but they would require a
finely tuned dependence of rotation velocity on metallicity to satisfy
a variety of observational constraints.

The $P\sim 10$ ms models investigated here can be viewed as a near lower limit
to the neutrino-driven wind contribution in the \citet{vlasov2017} models, as they have strong magnetic fields but minimal rotation, and they are already close to producing the observed levels of Sr, Y, and Zr.  Further reducing the predicted yields of these elements would require changing the electron fraction evolution in the cooling proto-NS models, by changing the ratio of the electron- and anti-electron neutrino fluxes in the first moments after successful SN explosion.

%%%%%%%%%%%%%%%%%%%%%%%%%%%%%%%%%%%%%%%%%%%

\section{Acknowledgments} 
We thank the referee, Friedrich-Karl Thielemann, for many precious and thought-provoking comments and suggestions, which improved the clarity and quality of our work. 
We thank Tuguldur Sukhbold for providing the set of stellar yields assumed in our chemical evolution models. 
We thank Anna Porredon and Sten Hasselquist for useful remarks. 
This work was supported by NSF grant AST-1909841.
F.V. acknowledges the support of a Fellowship from the Center for Cosmology and AstroParticle Physics at the Ohio State University. TAT is supported in part by NASA grant \#80NSSC20K0531. DW acknowledges support of the Hendricks Foundation at the Institute for Advanced Study. 

\section*{Data availability}
The data underlying this article will be shared on reasonable request to the corresponding author.

\end{document}